\documentclass[aps,prl,superscriptaddress,twocolumn,10pt]{revtex4}

\usepackage{amsfonts}
\usepackage{amsmath}
\usepackage{amssymb}
\usepackage[normalem]{ulem}
\usepackage{bm}
\usepackage{graphicx}
\usepackage{subfig}
\usepackage{lipsum}
\usepackage{array,multirow}
\usepackage[utf8x]{inputenc}

\def\opone{\leavevmode\hbox{\small1\kern-3.8pt\normalsize1}}
\usepackage{color}

\begin{document}
	
	\title{
		Proposal and Proof-of-Principle Demonstration of Fast-Switching Broadband Frequency-Shifting for a Frequency-Multiplexed Quantum Repeater}
	
	\author{Peng-Cheng Wang}
	\affiliation{QuTech and Kavli Institute of Nanoscience, Delft University of Technology, 2600 GA Delft, The Netherlands}
	\author{Oriol Pietx-Casas}
	\affiliation{QuTech and Kavli Institute of Nanoscience, Delft University of Technology, 2600 GA Delft, The Netherlands}
	\author{Mohsen Falamarzi Askarani}
	\affiliation{QuTech and Kavli Institute of Nanoscience, Delft University of Technology, 2600 GA Delft, The Netherlands}
	\author{Gustavo Castro do Amaral}
	\altaffiliation{Corresponding author: amaral@puc-rio.br}
	\affiliation{QuTech and Kavli Institute of Nanoscience, Delft University of Technology, 2600 GA Delft, The Netherlands}
	\affiliation{Center for Telecommunication Studies, Pontifical Catholic University of Rio de Janeiro,\\ 22451-900, Rio de Janeiro, Brazil}
	
	\begin{abstract}
		A proposal for fast-switching broadband frequency-shifting technology making use of frequency conversion in a nonlinear crystal is set forth, whereby the shifting is imparted to the converted photons by creating a bank of frequency-displaced pump modes that can be selected by a photonic switch and directed to the nonlinear crystal. Proof-of-principle results show that the expected frequency-shifting operation can be achieved. Even though the dimensions of the currently employed crystal and significant excess loss in the experimental setup prevented conversion of single-photon-level inputs, thorough experimental and theoretical analysis of the noise contribution allowed for the estimation of the system performance in an optimized scenario, where the expected signal-to-noise ratio for single-photon conversion and frequency-shifting can reach up to 25 dB with proper narrow-band filtering and state-of-the-art devices. The proposed frequency-shifting solution figures as a promising candidate for applications in frequency-multiplexed quantum repeater architectures with 25 dB output SNR (with 20\% conversion efficiency) and capacity for 16 channels spread around a 100 GHz spectral region.
		\end{abstract}
	
	\maketitle

\section{Introduction}

    The quantum internet is a network that links quantum processors through the distribution of entanglement. One feature of a working quantum internet is the possibility of implementing protocols such as cloud quantum computing, quantum-key-distribution (QKD), and dense coding \cite{Ekert(august1991)}, \cite{Benett(november1992)}.
    To be able to distribute entanglement through long-distance channels, optical photons are used, the main reason being the very low loss experienced in transmission through an optical fiber.
    
    \begin{figure*}[ht]
        \centering
        \includegraphics[width=0.95\linewidth]{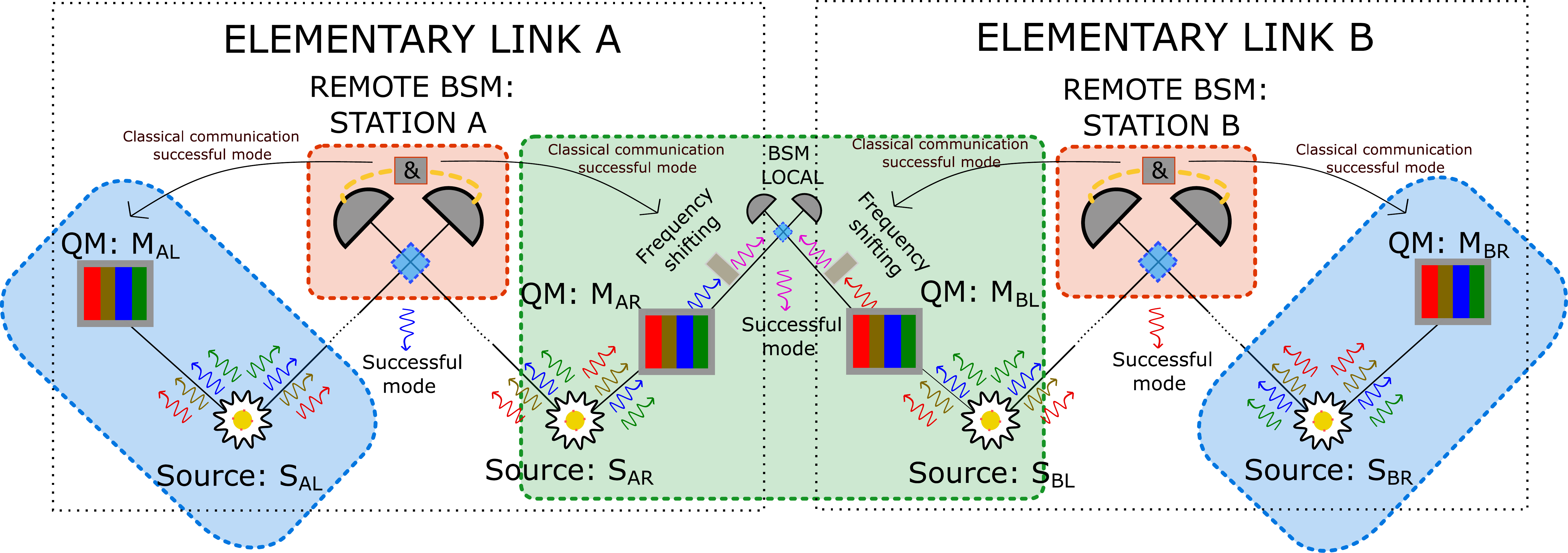}
         \caption{A representation of a frequency-multiplexed quantum repeater architecture. Source: Spectrally-Multiplexed Entangled-Photon-Pair Source; QM: Spectrally-Multiplexed Absorptive Quantum Memory; BSM: Spectrally-Multiplexed Bell-State Measurement. Shaded areas depict the same local region.}
         \label{fig:elementaryLinkFM}
    \end{figure*}
    
    The power loss experienced by an optical signal propagating in an optical fiber can be described by the Beer-Lambert law, $P(z) = P_0 e^{-\alpha z}$, expressed in terms of the attenuation coefficient $\alpha$, usually expressed in $dB/km$.  Although low, the propagation attenuation establishes a limit to the maximum distance between two nodes of the future quantum network in the range of hundreds of kilometers \cite{pirandola2017fundamental}. By making use of quantum repeater technology, the intrinsic direct-communication bound established by transmission loss can be circumvented. The goal is to break down long distances into smaller segments (elementary links), within which direct communication can be achieved. Each elementary link attempts to distribute entanglement independently and an entanglement swapping operation allows the links to be interconnected so that entanglement can be distributed to the end nodes of the network.
    
    The original proposal of a basic quantum repeater connects a string of imperfect entangled photon pairs by using a nested purification protocol \cite{Briegel(december1998)}. This proposition relies heavily on a technique called entanglement distillation, which is the purification of multiple long-distance imperfect shared-entangled pairs. 
    A different proposal of a quantum repeater architecture, the DLCZ \cite{Duan(november2001)} protocol, incorporates these purification protocols automatically into each of its elements. In this proposal, entanglement is created by making use of atomic ensembles as quantum memories and distributed through entanglement swapping. The quantum states are stored into the quantum memories until successful entanglement distribution over other elementary links is achieved, which, unfortunately, limits this protocol's entanglement distribution rates.
    An alternative approach that attempts to solve this issue was proposed by Sinclair et al. \cite{Sinclair(july2014)}. Similarly to the DLCZ protocol, concatenation of multiple elementary links is performed using quantum memories and two-photon interference, i.e., a projective measurement of the joint quantum state of the flying photons onto the Bell basis, a so-called Bell State Measurement (BSM). In a BSM, the recording of a specific detection pattern allows one to herald a successful Bell-state projection and, thus, successful entanglement swapping. The latter protocol features absorptive quantum memories and external entangled photon-pair sources (EPPSs) for entanglement generation.
    
    By harnessing the intrinsic properties of the quantum memories used in this quantum repeater architecture, mainly their broad spectrum, the concept of frequency multiplexing can be employed. Frequency multiplexing is a technique where the total available bandwidth in a communication channel is subdivided into a series of non-overlapping discrete frequency modes for increased communication throughput. In this case, the sources must be capable of emitting photon pairs at specific frequency modes and the BSM stations must be capable of individually processing all the frequency modes. Fig. \ref{fig:elementaryLinkFM} depicts a simplified diagram of such frequency multiplexed (FM) quantum repeater architecture. The FM photons are depicted in different colors, each color representing a distinct frequency mode. The following are the steps necessary for the operation of the elementary links; following this protocol, entanglement can, in principle, be distributed over long distances by concatenating any number of elementary links.

    \begin{itemize}
        \item Entangled photon pairs are generated by the source $\text{S}_{\text{AL}}$ and sent to both the remote BSM station A and the quantum memory $\text{M}_{\text{AL}}$. The same happens for source $\text{S}_{\text{AR}}$, remote BSM station A, and quantum memory $\text{M}_{\text{AR}}$.
        
        \item A BSM is performed in remote station A upon the synchronized arrival of photons from both sources. If the measurement is successful, then entanglement is swapped such that the quantum states stored inside $\text{M}_{\text{AL}}$, and $\text{M}_{\text{AR}}$ become entangled.
        
        \item Provided that the same happens in elementary link B, the states stored in quantum memories $\text{M}_{\text{AR}}$ (placed in the same location) and $\text{M}_{\text{BL}}$ can be retrieved.
        
        \item The information about which FM mode yielded a successful BSM result allows a frequency-shifting operation to take place such that the correct modes are directed to the local BSM.
        
        \item Conditioned on a successful measurement in the local BSM, the states stored in quantum  memories $\text{M}_{\text{AL}}$ and $\text{M}_{\text{BR}}$ become entangled.
    \end{itemize}
    
     Even though the propagation losses are constrained inside an elementary link, there are still limiting factors to the success of the entanglement swapping operation. Mainly, the photons that propagate to the remote BSM station experience attenuation and the success probability of the BSM is theoretically limited to 50\%, i.e., $\eta_{\text{swap}} = 0.5 \cdot e^{-2\alpha L}$ \cite{Calsamiglia(january2001)}, where the factor of 2 is related to the attenuation experienced by photons from different sides of the elementary link reaching the remote station. The availability of multiple spectral modes in the FM quantum repeater yields a higher probability of success for establishing entanglement per attempt across an elementary link. The probability that at least one out of $N$ frequency modes is successful can be written as 
     \begin{equation}
        P(\text{success}) = 1 - (1 - \eta_{\text{swap}})^N. 
    \end{equation}
    Thus, the success rate of at least one mode increases if one increases the amount of available frequency modes, thereby circumventing the limit imposed by the transmission loss and the BSM's efficiency.
    
    A crucial aspect of the entanglement swapping operation based on linear optics in a BSM is the indistinguishability between interfering photons. Following Fig. \ref{fig:elementaryLinkFM}, this is enforced in the remote BSM stations A and B by making sure that the emitted photons from all sources follow the exact same spectral distribution. It is, however, very unlikely that the spectral mode that has been successfully swapped in elementary link A is the same as for elementary link B. This creates the demand for a spectral mode mapper that maps any successfully swapped spectral mode to a previously agreed-upon spectral mode common to $M_{AR}$ and $M_{BL}$. An elegant solution is to communicate the successful mode to a frequency shifting device, a protocol also known as feed-forward spectral mode-mapping (FFSMM), which has been demonstrated in \cite{Puigibert(august2017)} by making use of an optical phase modulator driven by a fast linear voltage signal.
    
    \section{Frequency-Shifting Proposal}
    
    Here, a proposal is set forth whereby the frequency shifting is performed by means of a second-order non-linear optical process. Such a process enables mapping the wavelength of an input photon (signal) to a target wavelength (idler) mediated by a strong beam (pump) inside a nonlinear crystal. By shifting the center frequency of the pump beam, the idler's center frequency is also shifted, provided that the bandwidth of the conversion process is respected. Thus, the multiplexed spectrum retrieved from the quantum memory can be shifted in a such a way that the known successful spectral mode is the only one to be transmitted through a fixed narrow-band optical filter. Mathematically, let the successfully swapped spectral mode's center frequency be denoted by $\nu_{\text{sig}}+f_{\text{suc}}$ and the fixed optical filter's center frequency be denoted by $\nu_{\text{filter}}$, such that $\nu_{\text{pump}} = \nu_{\text{filter}}\pm \nu_{\text{sig}}$, where $\pm$ stands for \textit{difference} and \textit{sum} frequency generation, (DFG and SFG) respectively. Then, the required shift ($f_{\text{shift}}$) to be imparted to the pump is the one that allows for:
    \begin{equation}
        \left(\nu_{\text{pump}} \pm f_{\text{shift}}\right) \pm \left(\nu_{\text{sig}}\mp f_{\text{suc}}\right) = \nu_{\text{filter}}
    \label{eq:shift}
    \end{equation}
    
    As previously mentioned, such a FFSMM has been demonstrated by directly modulating the phase of the spectral modes retrieved from the quantum memory \cite{Puigibert(august2017)}. However, three main limitations are associated to this method: the insertion loss intrinsic to electro-optic phase modulators, which acts directly on the single photons; a bound on the maximum frequency shift that can be imparted to the optical signal, related to the frequency response of the phase modulator itself and from the ramp generation electronics; and the switching rate of the frequency shift, i.e., the minimum response time of the FFSMM for two distinct shift values $f_{\text{shift}}^i$ and $f_{\text{shift}}^j$ that need to be imparted to subsequently retrieved photons $i$ and $j$. Since, in the currently proposed scheme, the shift is indirectly imparted to the retrieved photons by correctly preparing the pump, most of these limitations can be lifted.
    
    \begin{figure}[ht]
        \centering
        \includegraphics[width=0.95\linewidth]{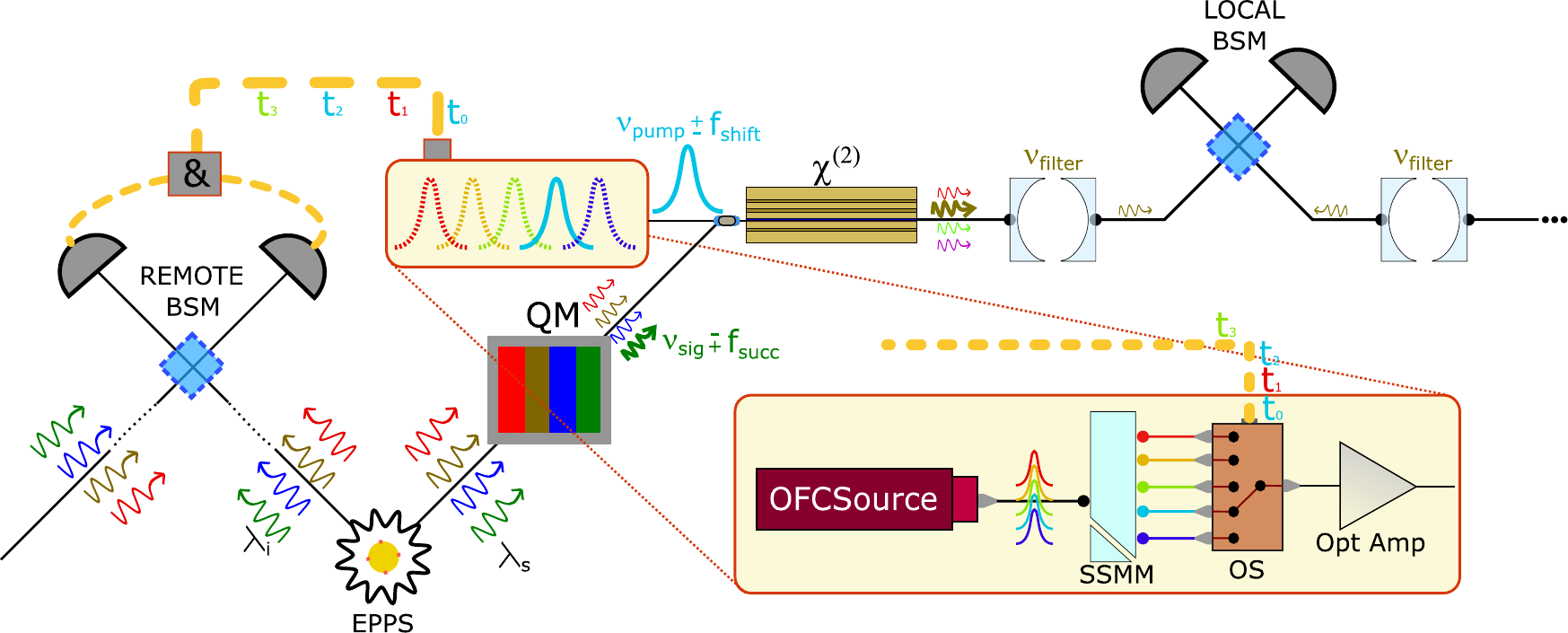}
         \caption{Proposal of an ideal frequency shifting setup, where the result from the remote BSM station is communicated to the pump preparation section so that the correct mode can be used for shifting. In the zoomed-in section, the pump preparation is composed of an OFC source, a SSMM, an optical switch (OS), and an optical amplifier.}
         \label{fig:proposalSetup}
    \end{figure}
    
    The pump preparation, highlighted in Fig. \ref{fig:proposalSetup}, is composed of an optical frequency comb (OFC) source, a spectral-to-spatial mode mapper (SSMM), and a Nx1 photonic switch. Although different techniques can be employed to create both the OFC \cite{sandoval_2019} and the SSMM \cite{bulkGrating}, here the proposal of \cite{Dou2012Generation} and a virtually-imaged phased-array (VIPA) \cite{1417049} are considered, respectively. In the former, the manipulation of the electrical signal that is used to drive a series of electro-optic modulators (EOM) allows for the generation of a comb with 300 GHz total bandwidth with overall 1.5 dB spectral flatness. Furthermore, the VIPA is an optical device that spatially separates light according to its spectral components using angular dispersion; it is considered here due to its broad bandwidth and high spectral resolution, where both become extremely important as the density of spectral modes increases. By connecting the OFC and SSMM, one produces an array of spectral modes demultiplexed in space. Each spectral mode can, then, be individually coupled to the inputs of a fast photonic switch \cite{Miao:14}. 
    
    Following the schematic depicted in Fig. \ref{fig:proposalSetup} the EPPSs emit photons within $N$ distinct frequency modes and the QMs store the quantum states for a fixed amount of time that corresponds to the round trip time to and from the remote BSM station, i.e., the time necessary for the photons to arrive at the remote BSM and the results to be communicated back. The results contain the information regarding which mode was successful, allowing for the frequency shifting setup to select the corresponding frequency shift according to Eq. \ref{eq:shift}.
    
    \section{Proof-of-Principle Results}
    
\begin{figure}[ht]
        \centering
        \includegraphics[width=0.95\linewidth]{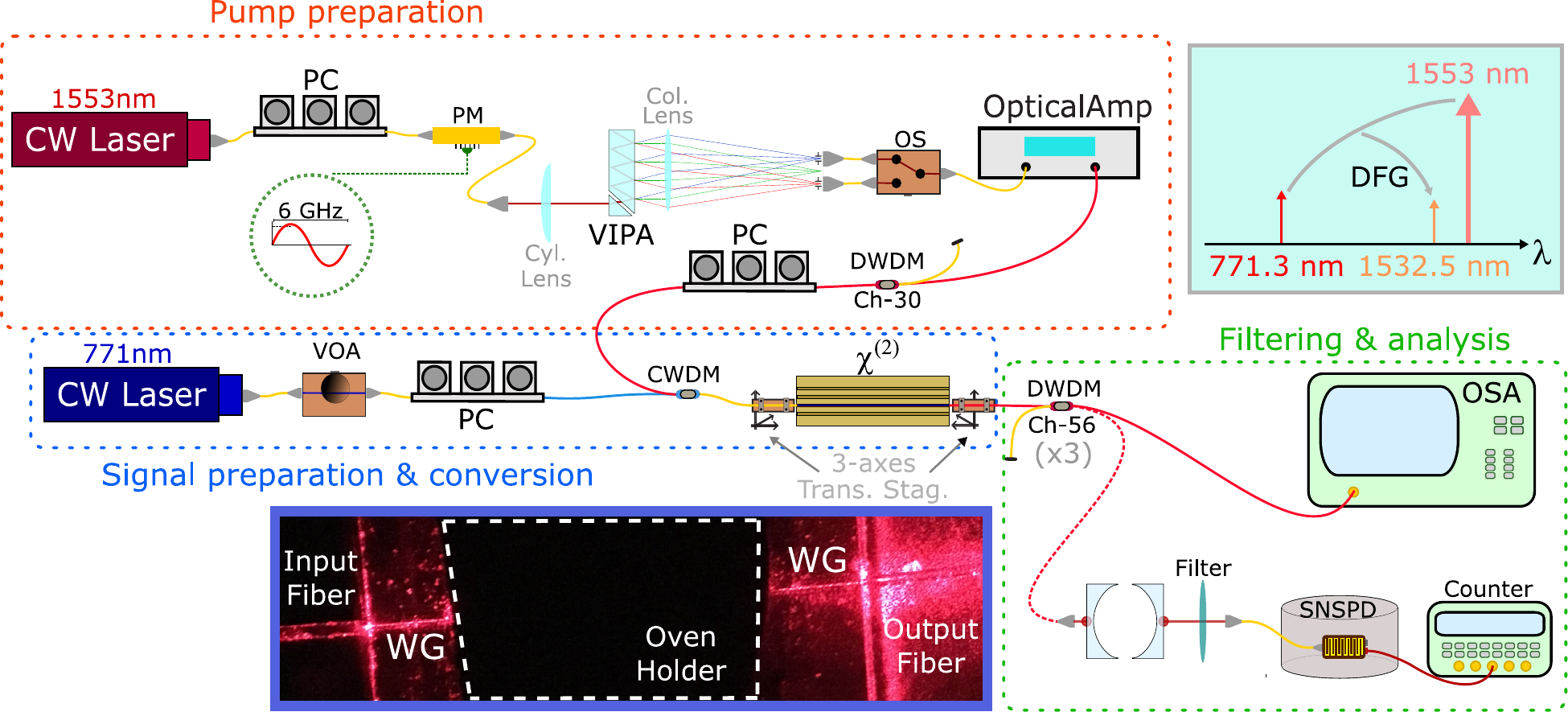}
         \caption{Schematic representation of experimental setup used for the proof-of-principle shifting results. The setup consists of three main sections: the pump preparation section; the signal preparation and conversion section; and the filtering and analysis section. In the upper inset, the wavelength pane, depicting the DFG process; in the bottom inset, a microscope picture of the fiber-coupling using a visible laser source.}
         \label{fig:expSetup}
    \end{figure}

    The experimental setup is depicted in Fig. \ref{fig:expSetup}, where $\lambda_{\text{sig}}$, $\lambda_{\text{filter}}$, and $\lambda_{\text{pump}}$ correspond to 771.3, 1532.5, and 1553 nm, respectively. In this case, it is more natural to refer to the wavelength of the optical beam instead of its frequency, as in the previous section. These will be used interchangeably and their relationship is given, as always, by $\lambda = \nu/c$. A frequency-stabilized ($\Delta \nu \leq 100$ KHz) laser source emitting at 1553 nm is directed to the pump preparation section, which consists of an EM driven by a 6 GHz sinusoidal signal (generating optical side-bands at $\nu_{\text{pump}}\pm6$ GHz), a VIPA matched to the frequency of the side-bands, and a 2x1 photonic switch. The output is directed to an optical amplifier capable of outputting up to 27 dBm of optical power. After filtering (DWDM channel 30), the pump signal was combined to the input (generated by an external-cavity solid-state laser) in a wavelength combiner and sent to a periodically-poled 0.5-cm-long LiNbO$_3$ crystal waveguide. Coupling into and out of the crystal waveguide is achieved through fibers held by 3-axes translation stages. The crystal is placed inside a temperature-controlled oven, which, at $T=40^o$C is phase-matched for DFG from 771.3 nm down to 1532.5 nm mediated by 1553 nm. Signal preparation involves polarization and intensity control with a manual polarization controller and a variable optical attenuator, respectively.
    
\begin{figure}[ht]
        \centering
        \includegraphics[width=0.95\linewidth]{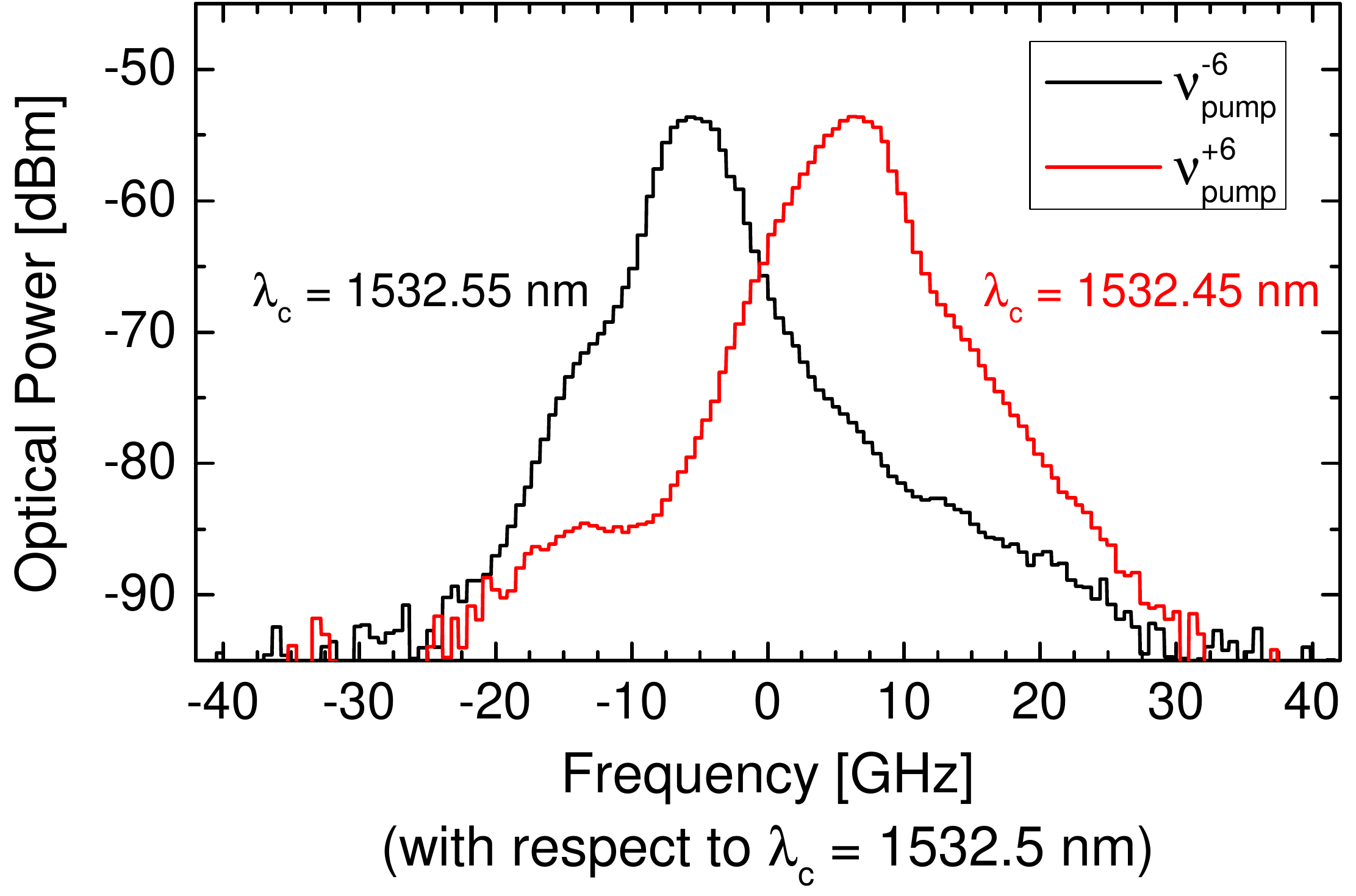}
         \caption{Output spectra measured in the optical spectrum analyzer (OSA) when the output of the optical switch is connected to either input that correspond to $\nu_{\text{pump}}^{\text{+6}}$ and $\nu_{\text{pump}}^{\text{-6}}$.  The spectral resolution of the OSA is set to 0.05 nm.}
         \label{fig:shiftChar}
    \end{figure}

    The output of the crystal waveguide is sent to a series of fibered optical filters centered at $\lambda_{\text{filter}}$ (channel 56 of the DWDM grid) and, then, to a free-space narrowband etalon cavity (FSR = 500 GHz, $\Delta \nu = 8$ GHz) also centered at $\lambda_{\text{filter}}$. Finally, the output beam is analyzed either in an optical spectral analyzer (OSA) or a single-photon detector (superconducting nanowire cooled down to 0.7 K in a cryostat exhibiting $\eta_{\text{det}}=50\%$, 2000 DCR, and 100 ns dead time). Characterization of the conversion and shifting achieved by acting on the photonic switch and selecting different sidebands in the pump preparation section can be found in Fig. \ref{fig:shiftChar}, where the measured spectra for the two configurations of the photonic switch are presented. As expected, the down-converted light is shifted by 12 GHz when the photonic switch is triggered, which can be directly verified from the measured data. For these results, $P_{\text{pump}} = 18$ dBm and $P_{\text{sig}} = -3$ dBm.
    
\section{Analysis of Practical Implementations}

    The results of Fig. \ref{fig:shiftChar} prove that the frequency-shifting mediated by frequency-conversion proposal is achievable. In this Section, the parameters of a practical implementation of this proposal are analyzed in terms of the presented experimental setup and the state-of-the-art. Unfortunately, the short length of the employed crystal, combined with an excess noise stemming from Raman scattering of the pump light inside the optical fiber (discussed below), prevent current single-photon-level demonstrations. 
    
    \begin{figure}[ht]
        \centering
        \includegraphics[width=0.95\linewidth]{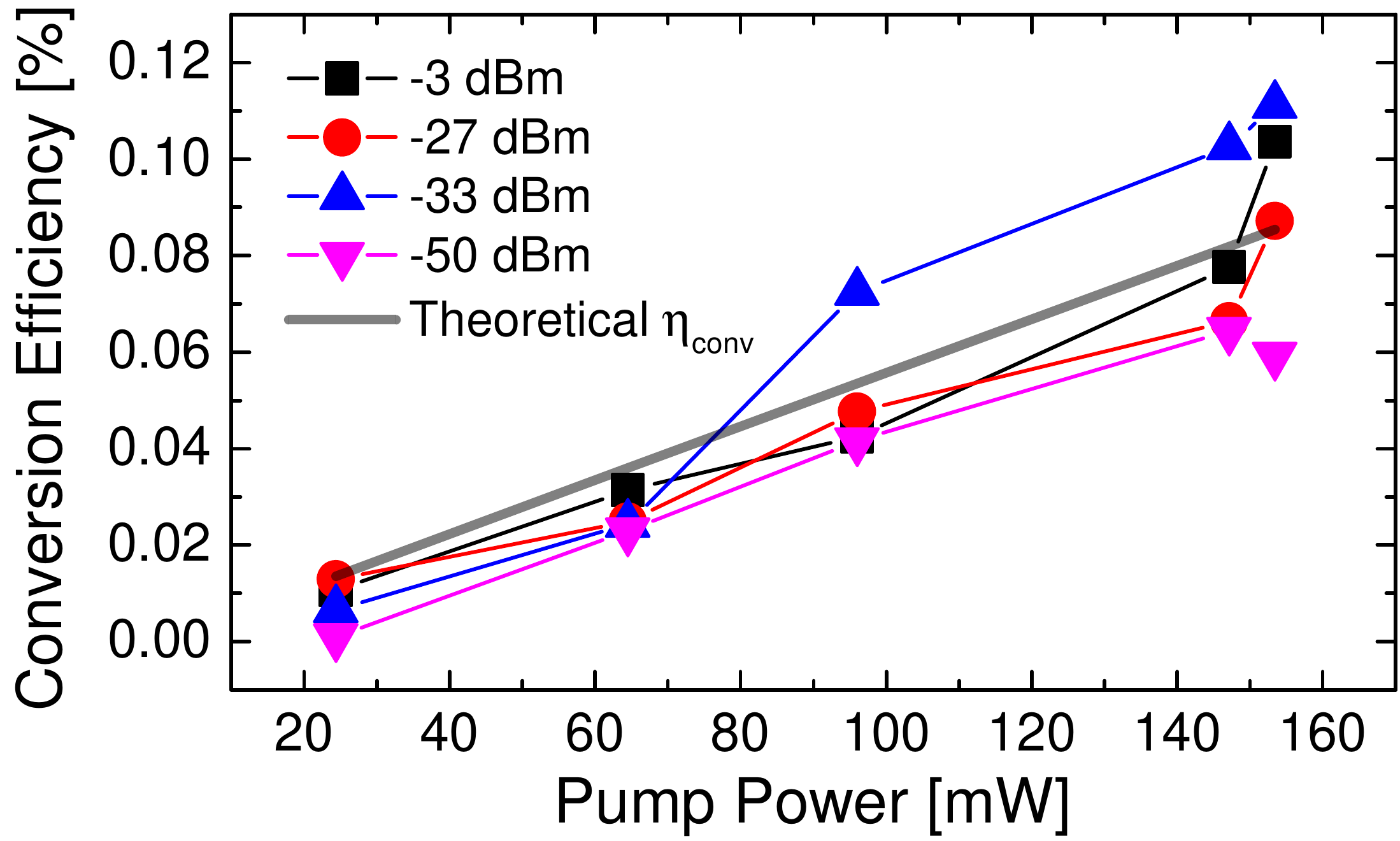}
        \caption{Theoretical conversion efficiency of the current experimental setup as a function of pump power supported by the experimental findings using different input signal powers.}
        \label{fig:convEff}
    \end{figure}
    
    As can be observed in Fig. \ref{fig:convEff}, the conversion efficiency of the crystal is limited to -30 dB at the maximum available pump power of 160 mW. The theoretical result is determined through:
    \begin{equation}
        \eta_{\text{conv}} = \sigma_{\text{WG}}^{\text{out}}\sin ^2\left(\sqrt{\sigma_{\text{LiNbO}_3}P_{\text{WG}}}L\right),
        \label{eq:convEff}
    \end{equation}
    where $P_{\text{WG}} = P_{\text{pump}}\sigma_{\text{WG}}^{\text{in}}$, $\sigma_{\text{WG}}^{\text{in}}$ $\sigma_{\text{WG}}^{\text{out}}$ are the measured input and output coupling efficiencies into/out of the crystal waveguide ($\sigma_{\text{WG}}^{\text{in}}=\sigma_{\text{WG}}^{\text{out}}\sim 0.3$), and $\sigma_{\text{LiNbO}_3}$ is a parameter of the crystal and equal to 0.48 W/cm$^2$ as specified by the manufacturer. This is confirmed by experimental results for various values of $P_{\text{pump}}$ with varying input signal powers ranging from -50 to -3 dBm. Under these conditions, it is possible to calculate the average spectral mode extinction ratio when the narrowband free-space cavity is centered at $\nu_{\text{filter}}^{+6}$($\nu_{\text{filter}}^{-6}$) and the optical switch is configured to output $\nu_{\text{pump}}^{-6}$($\nu_{\text{pump}}^{+6}$) to be 8dB.
    
\begin{figure}[ht]
        \centering
        \includegraphics[width=0.95\linewidth]{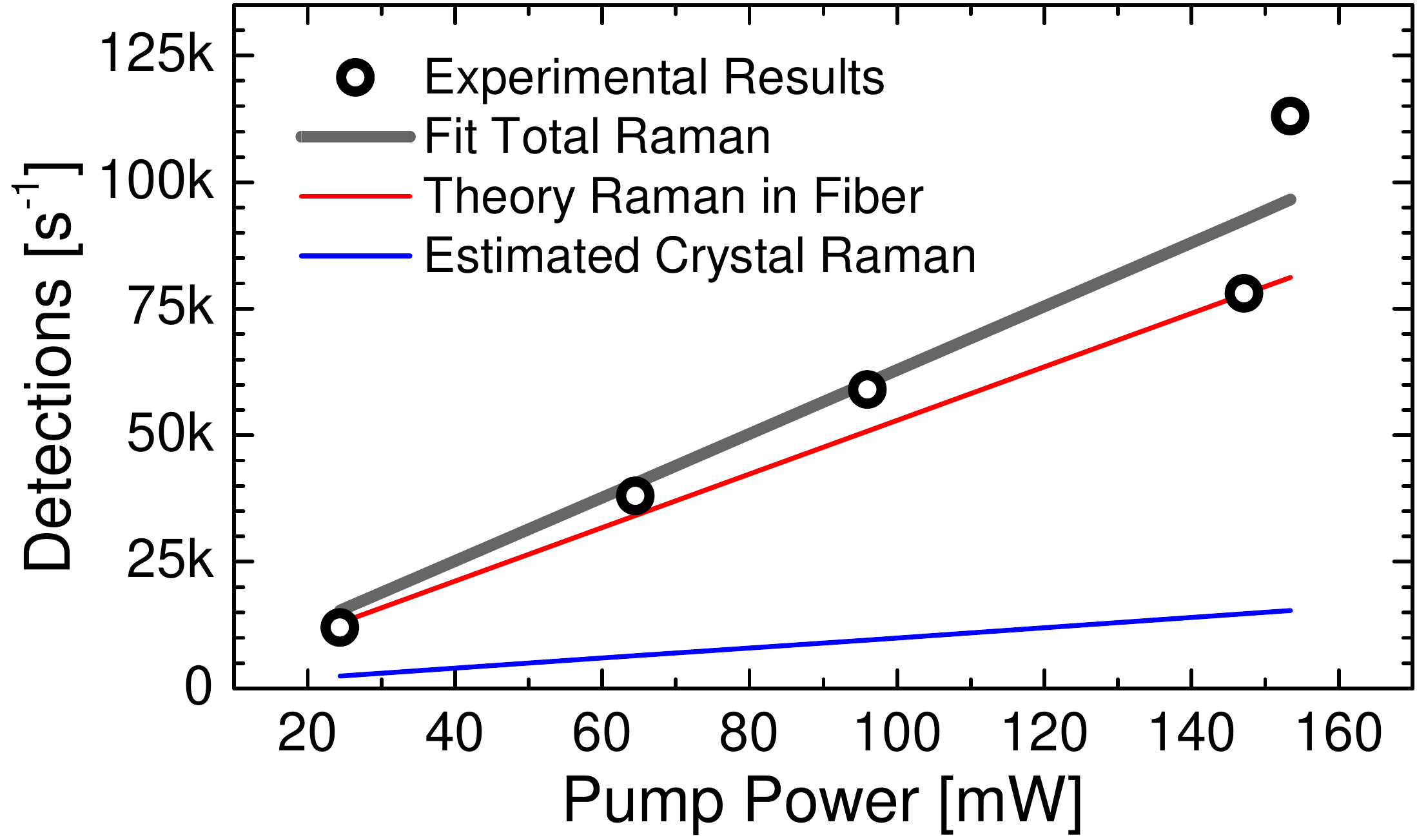}
        \caption{Estimation of the Raman scattering stemming from the pump propagating through the crystal waveguide (blue line), in detections per second. It corresponds to the subtraction between the experimentally determined overall noise in the system (black circles and dark grey line) and the theoretically determined Raman scattering stemming from the pump propagating through the optical fiber (red line).}
        \label{fig:ramCounts}
    \end{figure}

    By blocking the input light at 771.3 nm and varying the pump power, it is possible to determine the excess noise in the system. Simultaneously, the theoretical prediction of Raman scattering power due to the pump light propagating in the fiber can be determined through:
    \begin{equation}
        P_{\text{ram}} = P_{\text{in}}(0) \beta e^{-\alpha z},
        \label{eq:ramanScat}
    \end{equation}
    where $P_{\text{in}}(0)$ is the optical power propagating through the medium, z is the length of the medium, and $\beta$ and $\alpha$ are the (well-known for a fiber) spontaneous Raman coefficient and the attenuation coefficient of the medium, respectively \cite{Chapuran(october2009),Silva(july2014),smf28Thorlabs}. The analysis, depicted in Fig. \ref{fig:ramCounts}, allows one to estimate the Raman noise stemming from the propagation of the pump light through the crystal waveguide alone. This noise contribution is intrinsic to the frequency-conversion process in a non-linear medium and estimating it for future realizations is essential for a complete analysis. By employing a fit using Eq. \ref{eq:ramanScat}, $\beta_{\text{cryst}}$ and $\alpha_{\text{cryst}}$ of the LiNbO$_{3}$ crystal waveguide can be estimated.
    
\begin{figure}[ht]
        \centering
        \includegraphics[width=0.95\linewidth,trim={0 0 50pt 0},clip]{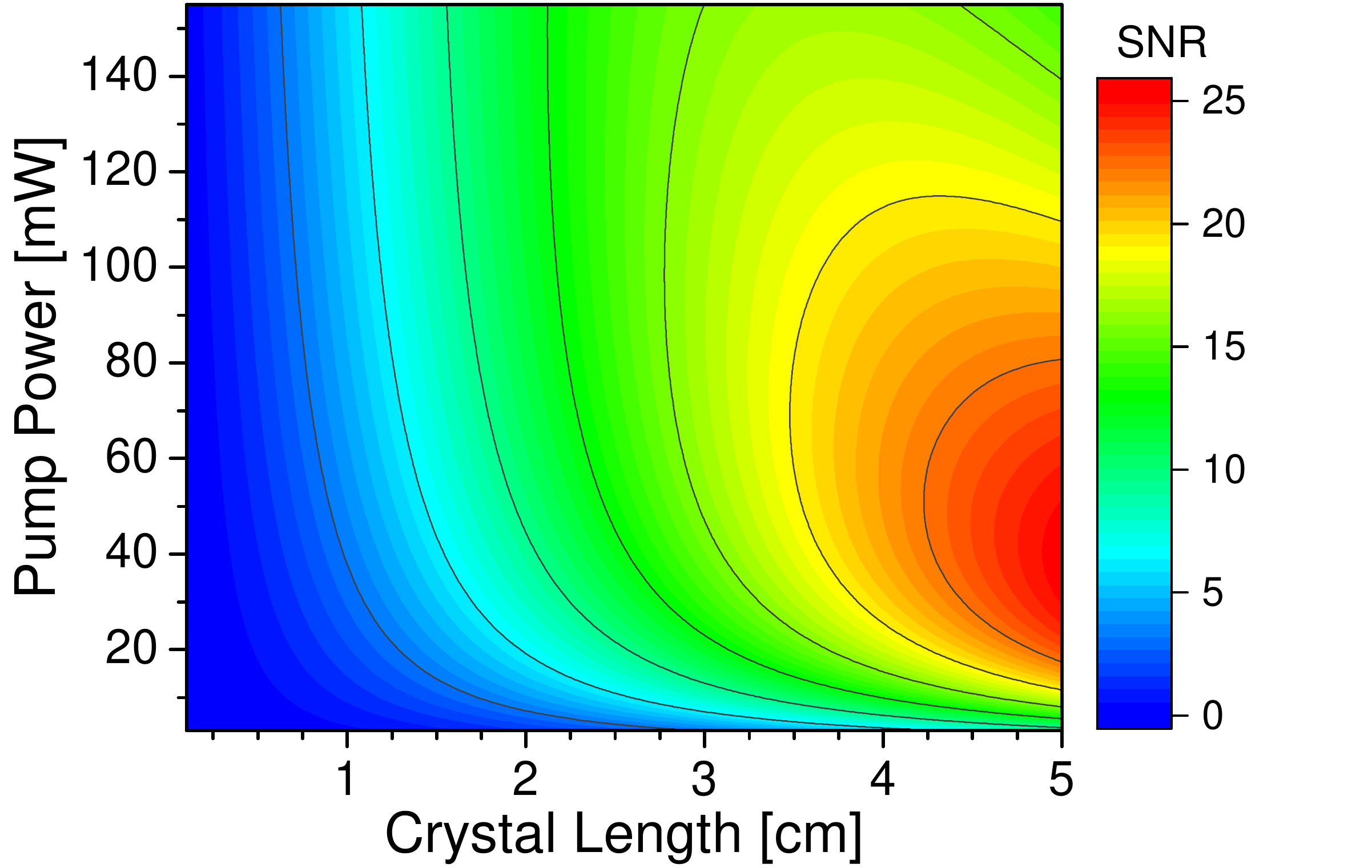}
        \caption{Signal-to-noise ratio at the output of the frequency-shifting setup as a function of pump power and crystal length for the implementation considering state-of-the-art devices. The maximum length of 5 cm for the crystal waveguide is stipulated based on values found in the literature.}
        \label{fig:expecSNR}
    \end{figure}

    For the expected signal-to-noise ratio (SNR) analysis, we consider a setup similar to the one in Fig. \ref{fig:expSetup}, but where the fiber-coupling into and out of the waveguide is replaced by free-space coupling, as in \cite{Maring(july2018)}. For that purpose, all fibered filters also need to be replaced by free-space filters, which is not a limiting issue. A narrowband filter ($\Delta \nu$ =250 MHz) is also considered, as also in \cite{Maring(july2018)}, to greatly suppress the unavoidable Raman scattering contribution originated in the crystal waveguide. Finally, detector parameters are considered as the ones measured in the current experiment. With these and the parameters previously determined, it is possible to estimate -- using Eqs. \ref{eq:convEff} and \ref{eq:ramanScat} -- the SNR of a practical frequency-shifting setup in the near future when the input is at the single-photon-level, i.e., $\mu = 0.1$. The results, as a function of pump power ($P_{\text{pump}}$) and crystal length ($L$), are depicted in Fig. \ref{fig:expecSNR}, where a 25 dB SNR is achievable for reasonable values of $P_{\text{pump}}$ and $L$, e.g., 60 mW and 5 cm, respectively.
    
    \subsection{Discussion}
    
    The proposed frequency-shifting solution has the potential to deliver 25 dB SNR conversion, with an internal conversion efficiency of 20\%; higher efficiencies (up to 25\%) can be achieved by sacrificing the SNR (down to 12 dB). Under these conditions, it is interesting to analyze the spectral and timing characteristics of the setup. The maximum shifting bandwidth is limited by two factors: the bandwidth of the OFC source; and the bandwidth of the SSMM. On one hand, in \cite{Dou2012Generation}, an OFC with total bandwidth of 300 GHz and a spectral flatness of 1.5dB has been reported. On the other hand, the bandwidth of the VIPA can be easily increased (by manipulating the VIPA width) to match this value, which is accompanied by a diminished achievable spectral resolution. This, in turn, translates into a smaller extinction ratio (ER) between spectral modes at the output of the SSMM, which, of course, impacts the ER after conversion.
    
    For comparison purposes, the currently employed VIPA, which offers a $\sim$60 GHz bandwidth with a spectral resolution of $\sim$0.6 GHz, was capable of delivering an 8 dB ER after conversion for spectral modes separated by 12 GHz. Optimized spatial mode filtering and fiber coupling could improve this up to 18 dB, as follows. The bandwidth of the VIPA, its spectral mode bandwidth (full-width at half maximum), and the cross-talk for a given spectral mode $j$ are given, respectively, by \cite{1417049}:
    \begin{equation}
    \begin{split}
        \Delta\nu_{\text{VIPA}} &= \frac{c}{2 t \cos\left(\theta_{\text{in}}\right)};\\
        \text{FWHM}_{\text{VIPA}} &= \frac{\Delta\nu_{\text{VIPA}}}{\pi} \frac{1-Rr}{\sqrt{Rr}};\\
        \text{CT}_{\text{VIPA}}^{j} &= \frac{\int_{\text{FWHM}} S\left(\omega\right)^{i=j}d\omega}{\int_{\text{FWHM}}\sum_i S\left(\omega\right)^{i\neq}d\omega},
    \end{split}
    \end{equation}
    where $c$ is the speed of light, $t=1.686$ mm is the VIPA's width, $R = 1$ and $r=0.95$ are the back and front reflectivities of the VIPA's interfaces, respectively, and $S\left(\omega\right)$ is the spectral power density of each individual mode at the focal plane. Considering the center frequency to be $\nu_{\text{pump}}$, and a total of 16 spectral modes, the physical parameters of the VIPA can be manufactured such that $\Delta\nu_{\text{VIPA}} = 98$ GHz, which yields $\text{FWHM}_{\text{VIPA}} = 1.2$ GHz, and $\text{CT}_{\text{VIPA}}^{j} = 18$ dB. A device based on bulk grating technology (BGT) exhibits lower crosstalk (25 dB), but offers less spectral flexibility \cite{kylia}. Therefore, a total of 16 spectral modes, covering a 100 GHz-wide spectral region withing the telecommunication C-band with reasonable crosstalk can be achieved.
    
    In terms of timing characteristics, the limitations are associated to the optical switch and optical amplifier. Fast optical switches with multiple input modes have recently been reported \cite{Cheng:18}, achieving ns-level switching times in a 16x16 configuration, which could be simplified into a 16x1 configuration, with low crosstalk between channels and almost lossless input coupling (due to the semiconductor optical amplifiers included in the design) \cite{Miao:14}. Low-noise high-output and broadband optical amplifiers are required in the proposed configuration in order to provide the necessary pump powers for high-efficiency conversion while maintaining a high output SNR. These values are within the region [40:160] mW according to Fig. \ref{fig:expecSNR}, i.e., necessary output powers in the order of 27 dBm. In \cite{keopsys}, EDFAs covering the telecommunication C-band and exhibiting 42 dBm maximum output power and 7.5 dB noise figure are reported. 
    
    Moreover, such amplifiers can operate with optical inputs as low as -20 dBm, so the power splitting due to the OFC generation and the subsequent coupling losses in both the SSMM and in the optical switch do not prevent operation, as follows. Total losses in the OFC setup amount to the splitting ratio (12 dB -- 16 spectral modes generated) and the insertion loss of the EOMs ($\sim$10dB). VIPA coupling losses have been reported as 3dB \cite{1417049} and the optical switches are almost lossless. Provided the employed laser source achieves an output power of 5 dBm, it will be enough to deliver -20 dBm at each spectral mode and, thus, to produce the required pump power for efficient conversion. The output of the amplifier can be further filtered in order to get rid of unwanted amplified spontaneous emission (ASE) by a series of DWDM filters ($\Delta\lambda = 0.8\text{nm} \implies \Delta\nu = 100\text{GHz}$), since the total bandwidth of the shifted pump modes (as previously determined) falls within this spectral interval. Table \ref{tab:summaryFigMer} presents a summary of the expected figures of merit of the proposed frequency-shifting solution making use of state-of-the-art technology.
    
    \begin{table}[!htb]
    \centering
    \caption{Expected Figures of Merit for the Proposed Frequency-Shifting Solution}
    \setlength\tabcolsep{5pt}
    \footnotesize\centering
    \smallskip 
    \begin{tabular}{c|c}
    Figure of Merit & Value\\
    \toprule
    Output SNR & 12 - 25 dB\\
    Conversion Efficiency & 20 - 25 \% \\
    Spectral Bandwidth & 100 GHz\\
    Number of Modes & 16\\
    Extinction Ratio & 18 dB\\
    Switching Time & $<$300 ns
    \end{tabular}
    \label{tab:summaryFigMer}
    \end{table}
    
    \section{Conclusion}
    
     Quantum repeater technology has the potential to interconnect quantum processing nodes for a future implementation of the quantum internet. In order to increase the throughput of entanglement distribution rate between nodes, spectral-multiplexing stands as a promising candidate. Its availability depends on the development of several building blocks, mainly spectral-multiplexed entangled-photon pair sources, quantum memories, Bell-State measurement stations and feed-forward spectral mode-mappers. In this work, a proposal for the implementation of a fast-switching broadband FFSMM has been set-forth with near-future estimated parameters that testify towards its implementation in the architecture of a spectrally-multiplexed quantum repeater. Proof-of-principle results of its shifting capabilities have been demonstrated and, while limited by current experimental limitations, allowed for the estimation of the setup's performance. Realization of a setup that exhibits the figures of merit estimated in this work is an on-going research focus.

\section{Acknowledgments}

The authors thank W. Tittel for fruitful discussions and financial support. The authors acknowledge funding through the Netherlands Organization for Scientific Research (NWO).

\end{document}